\documentclass[12pt]{article}

\usepackage{graphicx}
\usepackage{amsmath}
\usepackage[super,comma]{natbib}
\begin{document}

\begin{center}
{\bf Why does high pressure destroy co-non-solvency of PNIPAm in aqueous methanol?}
\end{center}

\begin{center}

{Tiago E. de Oliveira and Paulo A. Netz}

{\it Max-Planck Institut f\"ur Polymerforschung, Ackermannweg 10, 55128 Mainz Germany}
{\it Universidade Federal do Rio Grande do Sul, Porto Alegre, Brazil}

\vspace*{0.2in}

{Debashish Mukherji{\footnote {mukherji@mpip-mainz.mpg.de}} and Kurt Kremer{\footnote {kremer@mpip-mainz.mpg.de}}}

{\it Max-Planck Institut f\"ur Polymerforschung, Ackermannweg 10, 55128 Mainz Germany} 

\end{center}

\vspace*{0.2in}

{\bf Abstract:}
It is well known that poly(N-isopropylacrylamide) (PNIPAm) exhibits an interesting, yet puzzling, phenomenon of co-non-solvency.
Co-non-solvency occurs when two competing good solvents for PNIPAm, such as water and alcohol,
are mixed together. As a result, the same PNIPAm collapses within intermediate mixing ratios.
This complex conformational transition is driven by preferential binding of methanol with PNIPAm.
Interestingly, co-non-solvency can be destroyed when applying high hydrostatic pressures.
In this work, using a large scale molecular dynamics simulation employing high pressures,
we propose a microscopic picture behind the suppression of co-non-solvency phenomenon.
Based on thermodynamic and structural analysis, our results suggest that the preferential binding of methanol with
PNIPAm gets partially lost at high pressures, making the background fluid reasonably homogeneous for the polymer. This
is consistent with the hypothesis that the co-non-solvency phenomenon is driven by preferential binding and is not based on depletion effects.

\pagebreak

\section{Introduction}

Poly(N-isopropylacrylamide) (PNIPAm) is a so called smart polymer that responds to a wide range of external stimuli, such as
temperature, cosolvents, ionic strengths, and pressures. One of the most fascinating and puzzling phenomenon of
PNIPAm is its ability to exhibit co-non-solvency \cite{schild91mac,zhang01prl,walter12jpcb,koj12jpsb,tanaka08prl,mukherji13mac,mukherji14natcom}.
When a sample of PNIPAm is dissolved in mixtures of water and alcohol under ambient conditions, it collapses when the
composition of solvent mixtures are between $5-40\%$ of alcohol concentration \cite{schild91mac,zhang01prl,walter12jpcb,koj12jpsb}. Understanding this complex
structural transition is not only scientifically challenging \cite{mukherji13mac,mukherji14natcom}, but also has
a wide variety of applicabilities that range from physics to biology \cite{cohen10natmat,ward11poly,sissi14natcom}.
In this context, it has been recently shown that the co-non-solvency can only be explained by the preferential binding
of one of the cosolvent components with the polymer. In other words, the competitive displacement of
cosolvent components play a significant role in describing co-non-solvency \cite{mukherji14natcom,mukherji15jcp}.
It was suggested that when a very small amount of the better cosolvent is added into the dilute aqueous polymer solution,
these better cosolvents bind two monomers potentially far along the backbone forming segmental loops.
This loop formation initiates the process leading to a final well collapsed structure of the polymer. Interestingly, this preferential
cosolvent binding can also explain the reopening of the polymers at high cosolvent concentrations by the complete decoration of polymer with
cosolvents \cite{mukherji14natcom,mukherji15jcp}.

Another surprising phenomenon of PNIPAm is when they are exposed to high hydrostatic pressures.
It was experimentally observed that when a collapsed PNIPAm between $5-40\%$ of
alcohol concentration is put under high hydrostatic pressures at 298 K, co-non-solvency gets completely destroyed. As a consequence,
a PNIPAm chain only remains in the expanded coil state, irrespective of the water-methanol mixing concentrations \cite{hofmann14pol}.
The present work is the first attempt to give a detailed microscopic picture of this interesting pressure induced reopening of PNIPAm
under co-non-solvency condition. We use large scale molecular dynamics simulations to study the
conformational transition of PNIPAm in aqueous methanol employing high hydrostatic pressures. We perform thermodynamic and
structural analysis to propose a microscopic origin of this high pressure effect.

The remainder of the paper is organized as follows: in section \ref{sec:metho} we briefly state the methodology for simulations and
section \ref{sec:result} presents results and discussion. Finally we draw our conclusions in section \ref{sec:conclu}.

\section{Simulation Method and Model}
\label{sec:metho}

In this study we employ all atom molecular dynamics simulations using GROMACS package \cite{gro}.
We use the Gromos96 force field \cite{groms} for methanol, the SPC/E water model \cite{spce}
and the force field parameters for PNIPAm are taken from Ref.~\cite{walter12jpcb}.
The temperature is set to 298 K using a Berendsen thermostat with a coupling constant 0.1 ps.
The time step for the simulations is chosen as 1fs. Unless stated otherwise results are shown for the ambient
and 500 MPa pressures. However, in some cases, we have also performed simulations at 100 MPa and 200 MPa
to systematically test the pressure effects. It should be noted that the all atom
force field used here has reasonably good transferability over a wide range of pressures and temperatures \cite{per01jpca}.
The pressure coupling is done using a Berendsen barostat\cite{berend} with a coupling time of 0.5 ps.
The electrostatics are treated using Particle Mesh Ewald \cite{pme}. The interaction cutoff is chosen as 1.4nm.

We use a PNIPAm chain of length $N = 32$ solvated in a simulation box consisting of $2 \times 10^4$ solvent molecules
at 25\% methanol molar concentration $x_m$, i.e. $0.5\times 10^4$ methanol and $1.5\times 10^4$ water molecules, respectively.
In some cases, we have also performed simulations over full concentration range of methanol, ranging from pure water $x_m = 0.0$
to pure methanol $x_m = 1.0$.
This system size is large enough to maintain solvent equilibrium between the local region within the
vicinity of polymer and the bulk aqueous methanol solution. Note that maintaining solvent equilibrium in molecular simulations is a
paramount task, which is most severe when the polymer collapse and expansion is driven by
strong local concentration fluctuations of different solvent components.
This can either be achieved by using a grand-canonical-like approach \cite{mukherji13mac} or
by using a large simulation box \cite{mukherji12jctclet}. Mid-sized simulation domains are prone to system size effects and, therefore,
may lead to unphysical structural fluctuations.
Every initial configuration is equilibrated for 50ns under ambient
pressure. The production runs are performed for 450 ns at 298 K and varying pressures.
During the production run observables such as end-to-end distance $R_{ee}$, pair distribution
function ${\rm g}_{ij}(r)$, Kirkwood-Buff integral $G_{ij}$ and potential of mean force $V_{\rm PMF} (r)$ are calculated.
The time scale of simulation used here is approximately one order of magnitude larger than the conformational relaxation time of a PNIPAm chain,
which is estimated by calculating the end-to-end autocorrelation function $\left<R_{ee}(t)\cdot R_{ee}(0)\right>$.

\section{Results and Discussions}
\label{sec:result}

\subsection{Polymer conformation under high pressures}

\begin{figure}[ptb]
\begin{center}
\includegraphics[width=0.49\textwidth,angle=0]{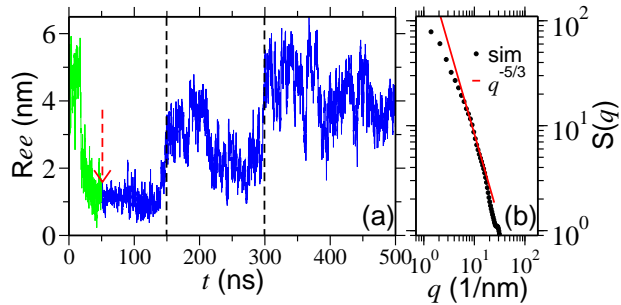}
\end{center}
\caption{Part (a) shows the time evolution of polymer end-to-end distance $R_{ee}$. The results are shown for a
chain length $N_l = 32$ and at a methanol concentration of 25\%. Initial equilibration starts with a
completely extended configuration of PNIPAm at a temperature of 298 K and ambient pressure (represented by green curve). A pressure of 500 MPa
pressure is employed beginning at 50ns (represented by the red arrow). Two vertical dashed lines are drawn to present different
time regimes during polymer reopening. Between $50~{\rm ns} < t <150~{\rm ns}$ the polymer remains fully collapsed, for $150~{\rm ns} < t < 300~{\rm ns}$
the end loops get open and finally polymer completely opens up for $t > 300~{\rm ns}$. Part (b) presents the static structure factor $S(q)$
of a PNIPAm backbone for $t > 300~{\rm ns}$. Note that for the calculation of $S(q)$ only alkane backbone was considered.
\label{fig:ree_t}}
\end{figure}

\begin{figure}[h!]
\begin{center}
\includegraphics[width=0.28\textwidth,angle=0]{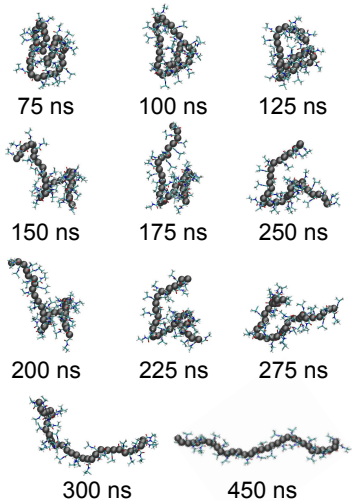}
\end{center}
\caption{Sequence of snapshots for a PNIPAm chain of length $N_l = 32$ at different times as measured during the
simulations. To better represent the polymer conformation, we render alkane backbone with spheres.
\label{fig:mov}}
\end{figure}

We start our discussion by presenting the central result of this paper, which is the structure of polymer at
high pressures. The initial configurations are generated by performing a simulation starting from a completely extended PNIPAm
structure at 298 K temperature and ambient pressure. In Fig.~\ref{fig:ree_t}(a) the green curve (for $t < 50~{\rm ns}$) presents the time evolution of polymer end-to-end
distance $R_{ee}$ during equilibration.
The structure collapses within 25 ns of MD run. Then we further monitor the collapsed structure for
another 25 ns to identify any unphysical fluctuations, which showed a rather stable collapsed conformation.
The last frame of this initially equilibrated sample was used for the production runs
under high pressures. The blue curve in Fig.~\ref{fig:ree_t}(a) presents time evolution of $R_{ee}$ at 500 MPa calculated over a 450 ns simulation trajectory.
It can be appreciated that the polymer remains within a completely globular state for almost $100~{\rm ns}$, with a
distinctly prominent stable polymer loop (see simulation snapshots in Fig.~\ref{fig:mov}).
The first expansion occurs at around $150~{\rm ns}$ when the end loop opens up. The complete opening of polymer chain
occurs for $t > 300~{\rm ns}$. A sequence of simulation snapshots is presented in Fig.~\ref{fig:mov}.
Thus our simulations could correctly capture the features observed in the high pressure experiments \cite{hofmann14pol}.

Furthermore, to confirm that we are indeed getting a well extended structure at 500 MPa, we look into the scaling law of static structure factor for
a PNIPAm chain at 500 MPa, which should support a scaling law $S(q) \sim q^{-1/\nu}$ with $\nu = 3/5$ being the Flory exponent \cite{degennesbook,desclobook}.
In Fig.~\ref{fig:ree_t}(b) we show $S(q)$ for a PNIPAm chain at 500 MPa and calculated from the MD trajectory for $t > 300~ns$.
Indeed, the data in the range $4~{\rm nm}^{-1} < q < 20~{\rm nm}^{-1}$ can be reasonably well described by a scaling exponent $\nu = 5/3$
known from the self avoiding random walk \cite{degennesbook,desclobook}. This range falls within the length scale of $1.6$ nm and $0.4$ nm.
Considering that the gyration radius $R_g \sim 1.7$ nm, the observed length scale is satisfactory.
Moreover, it should also be mentioned that ideally a good estimate of $S(q)$ requires long chains and here we are simulating a rather short chain
of $N_l = 32$ (or approximately 10 persistence lengths). Therefore, while the data in Fig.~\ref{fig:ree_t}(b) is certainly not good enough to derive an
aparent exponent, it is reasonable to clearly mark an extended chain.

\begin{figure}[ptb]
\begin{center}
\includegraphics[width=0.43\textwidth,angle=0]{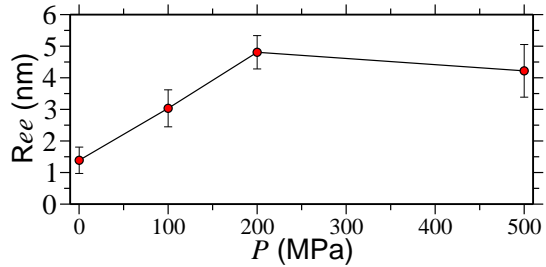}
\end{center}
\caption{Polymer end-to-end distance $R_{ee}$ as a function of applied pressure $P$ for a PNIPAm chain of length $N_l = 32$ and
at a temperature of 298 K.
\label{fig:ree_press}}
\end{figure}

Here, we also want to comment on the range of pressures used here and in the experiments \cite{hofmann14pol}. It should be
noted that a pressure of upto 200 MPa was used in Ref. \cite{hofmann14pol}. However, thus far, we have only presented results for 500 MPa.
Therefore, in Fig.~\ref{fig:ree_press} we show a systematic dependence of $R_{ee}$ on pressure. It can be appreciated that the
polymer reaches a fully extended state (represented by $R_{ee} \sim 4.5$) at $P \ge 200~{\rm MPa}$. This gives a very good
comparison with the experimental results. For $P = 100~{\rm MPa}$, however, we find a semi-collapsed structure
(with  $R_{ee} \sim 3.0$) for up to 450ns, the typical simulation time scale investigated here.

The observed prominent loops (see Fig.~\ref{fig:mov}) in our all atom simulations is reminiscent of the proposed mechanism
of polymer collapse transition in mixed good solvents \cite{mukherji14natcom}. It is known that the loops are
formed because of the bridging methanol molecules that can bind two distinctly far monomers along the
backbone \cite{mukherji14natcom}. Therefore, if the bridging is getting destroyed at high pressures, then there
must also be a disruption of methanol-polymer interaction to facilitate the opening of a PNIPAm chain. Therefore,
to establish a microscopic picture of the high pressure effects, we first look into the structure of the water and methanol within the solvation
volume of the polymer.

\subsection{Coordination and excess coordination numbers}

\begin{figure}[ptb]
\begin{center}
\includegraphics[width=0.37\textwidth,angle=0]{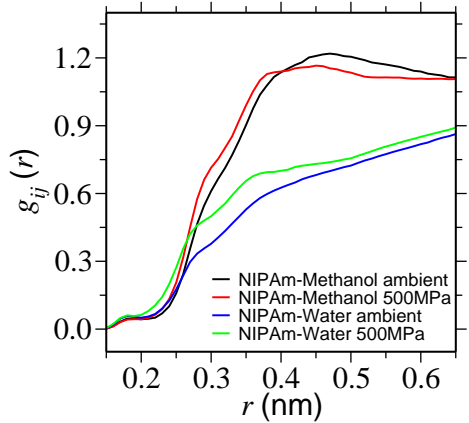}
\end{center}
\caption{Radial distribution function ${\rm g}_{ij}(r)$ showing NIPAm-methanol and NIPAm-water coordination
for two different pressures. Simulations are performed at a temperature of 298 K.
\label{fig:g_of_r}}
\end{figure}

In this section we perform structural analysis of the polymer solution.
For this purpose we calculate the radial distribution function ${\rm g}_{ij}(r)$ between
solution components. To obtain better convergence in ${\rm g}_{ij}(r)$, we have simulated a single monomer of
PNIPAm (represented as NIPAm) at a 25\% methanol-water mixture.
In Fig.~\ref{fig:g_of_r} we present NIPAm-methanol and NIPAm-water ${\rm g}_{ij}(r)$ for two
different pressures.
It is aparent from the plot that - while methanol coordination reduces partially within the first solvation shell (at around 0.5nm),
the coordination of water increases. This suggests that the methanol is getting partially replaced by water within the solvation shell
of the PNIPAm.

Furthermore, in table~\ref{tab:num} we present an estimate of the change in coordination number between
NIPAm and bulk solution components.
\begin{table}[h!]
\caption{A table listing various solute-solvent pairs with their respective coordination calculated using
$n = 4\pi\int_0^{0.5} g_{ij}(r)r^2 dr$, bulk solution number density of solution components $\rho$ and the
coordination numbers $n\rho$.}
\begin{center}
\begin{tabular}{|c|c|c|c|c|}
\hline
Pairs at different pressures & $n$ (${\rm nm}^3$) & $\rho$ (${\rm nm}^{-3}$)& $n\rho$ \\\hline
\hline
NIPAm-Methanol ambient &0.4718  &  6.7749 &  3.1964 \\
NIPAm-Methanol 500 MPa &0.4758  &  7.8068 &  3.7145 \\
NIPAm-Water ambient &0.2352  & 20.3248 &  4.7804 \\
NIPAm-Water 500 MPa &0.3123  & 23.4204 &  7.3142 \\
\hline
\end{tabular}  \label{tab:num}
\end{center}
\end{table}
It can be appreciated that with increasing pressure the coordination number of NIPAm-methanol only increases by about 16\%,
whereas NIPAm-water increase by 54\%. This suggests that the water is replacing methanol in the solvation shell, making the
background fluid more homogeneous for the polymers. This is consistent with the expanded structure of the polymer.

The density of the system increases about 15\%
when the system goes from ambient pressure to 500 MPa. It is known that this increase in density
leads to a substantial increase of the average coordination number of water \cite{net02},
and also to an increase in the diffusion coefficient at low temperatures \cite{net04}, but at high temperatures the
effect of the pressure on the diffusion coefficient is the opposite. Indeed, when the high pressure is applied,
the diffusion coefficient of water and methanol (data not shown) decrease by about 40\% and
50\%, respectively. Thus suggesting that the pressure-induced replacement of methanol with water has a thermodynamic
rather than a kinetic origin.

\begin{figure}[ptb]
\begin{center}
\includegraphics[width=0.46\textwidth,angle=0]{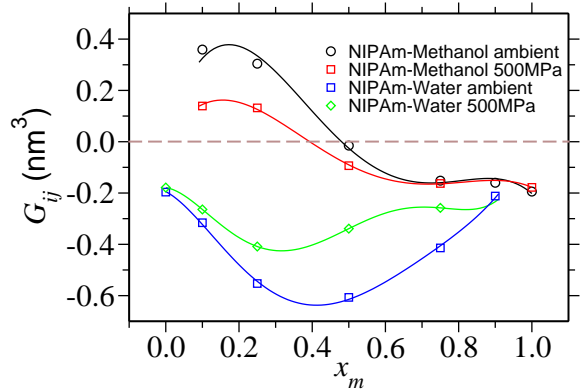}
\end{center}
\caption{Kirkwood-Buff integral $G_{ij}$ showing NIPAm-methanol $G_{pm}$ and NIPAm-water $G_{pw}$ excess coordination as a function of
methanol molar fraction $x_m$. Lines are the polynomial fits to the data that are drawn to guide the eye.
For pure solvent at $x_c = 0.0$ and pure cosolvent at $x_c = 1.0$, individual
coordinations $G_{pm}$ and $G_{pw}$ are undefined, respectively. Horizontal dashed line is
drawn to show $G_{ij} = 0$. The data corresponding to the ambient pressure is taken from Ref.~\cite{mukherji13mac}.
\label{fig:kbi}}
\end{figure}

A theory that perhaps best connects the relative intermolecular affinity and the solution thermodynamics is the
fluctuation theory of Kirkwood and Buff (KB) \cite{kb51jcp}. KB theory connects ${\rm g}_{ij}(r)$
to thermodynamic properties of solutions using the ``so called" KB integrals or
excess coordinations,
\begin{equation}
{\rm G}_{ij} = 4\pi \int_0^\infty \left[ {\rm g}_{ij}(r) - 1\right] r^2 dr.
\label{eq:kbi}
\end{equation}
In Fig.~\ref{fig:kbi} we summarize NIPAm-methanol $G_{pm}$ and NIPAm-water $G_{pw}$ excess coordination over
full molar concentration range of methanol $x_m$.
Ideally ${\rm G}_{ij}$ should be taken from the plateau at $r \to \infty$. Moreover, we estimate ${\rm G}_{ij}$ values by taking
averages between $0.9~{\rm nm} < r < 1.5~{\rm nm}$. Note that the typical correlation lengths in these systems are of the
order of $1.5~{\rm nm}$. It can be seen that - in comparison to NIPAm-water excess coordination, NIPAm-methanol still shows preferentiability even at
500 MPa. However, it is reduced by a factor of two. It is interesting to observe that the polymer opens up even when there remains
preferentiability. In this context, it is still important to mention that the fully collapsed structure needs a certain fraction of methanol
molecules within the solvation volume. Reduction in this fraction may not lead to a well collapsed conformation.
Instead, occasionally, one expects to observe a fluctuation in the extended polymer conformations, where instantaneous bridging may occur
(forming loops) due to a small fraction of methanol molecules within the solvation shell of PNIPAm.

To better quantify this reduced preferentiability one can translate the information presented in Fig.~\ref{fig:kbi} into chemical potential
of PNIPAm $\mu_p$, which can be calculated using \cite{rosgen05bio},
\begin{equation}
\frac{1}{k_{\rm B}T} \left(\frac {\partial {\mu}_{p}}{\partial {\rho}_m}\right)_{p,T} = \frac {{\rm G}_{pw} -{\rm G}_{pm}}
{1- {\rho}_m {\left({\rm G}_{mw} -{\rm G}_{mm}\right)}},
\label{eq:chempot}
\end{equation}
where $\rho_m$ is the methanol number density and $k_{\rm B}$ is the Boltzmann constant.
\begin{figure}[ptb]
\begin{center}
\includegraphics[width=0.46\textwidth,angle=0]{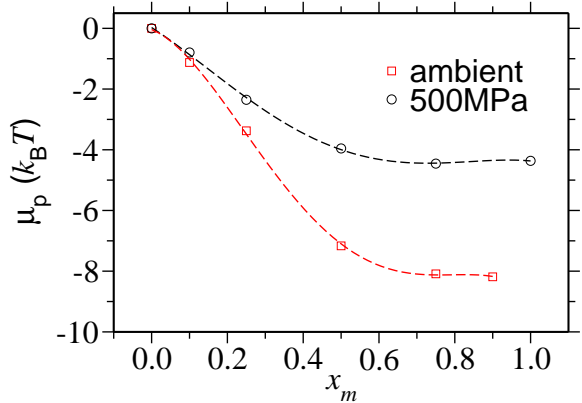}
\end{center}
\caption{Chemical potential shift per monomer $\mu_p/N_{l}$ as a function of methanol mole fraction ${x_m}$ for
two different pressures. The $\mu_p$ is calculated by integrating the data obtained from Eq.~\ref{eq:chempot}.
The data corresponding to the ambient pressure is taken from Ref.\cite{mukherji13mac}.
\label{fig:chem_pot}}
\end{figure}
In Fig.~\ref{fig:chem_pot} we show $\mu_p$ as a function of ${x_m}$ for different $N_l$'s, calculated by integrating
Eq.~\ref{eq:chempot}. For 500 MPa, it can be appreciated that the difference in $\mu_p$ between NIPAm in pure methanol (or $x_m = 1.0$)
and NIPAm in pure water (or $x_m = 0.0$) is reduced to $4k_{\rm B}T$, which is otherwise $8k_{\rm B}T$ under the ambient conditions.
Thus clearly indicating that by adding methanol molecules into the solution, the solvent quality is not getting as better as in the case
of ambient pressure. Note that the methanol driven collapse of PNIPAm under ambient condition occurs when the solvent quality remains good or
even gets increasingly better \cite{mukherji13mac,mukherji14natcom} and that this assymetry should be of the order of $8 - 10k_{\rm B}T$.
To further investigate the thermodynamic origin of this reduced preferentiability we also calculate potential of mean force in the next section.

It is yet important to mention that the polymer collapse can either be initiated by: (a) the bridging and looping
scenario presented earlier \cite{mukherji13mac} or (b) the depletion effects \cite{lekerbook}. Our arguments of polymer collapse-swelling transition is based on
the scenario (a). However, it could also be argued that the depletion effects \cite{lekerbook}, that are responsible for
polymer collapse under the poor solvent conditions, may be a factor behind PNIPAm collapse in aqueous methanol mixtures under ambient pressure.
However, it should be noted that when two competing good solvents are mixed together, such that the dissolved polymer
collapses within the intermediate mixing ratios, the collapse happens when the solvent quality remains good or even gets
increasingly better by the addition of better cosolvent (in this case methanol) \cite{mukherji13mac}.
This makes the polymer conformation decoupled from the solvent quality. Therefore, precluding any explanation
based on the depletion effects that can ``only" explain poor solvent collapse.
Furthermore, the depletion induced attractions can only be enhanced when increasing density. Note that for 500 MPa pressure bulk solution
density increases by $15\%$.
Therefore, if the pure depletion effects were the microscopic origin of co-non-solvency, PNIPAm would never open under the influence of
higher pressures. The same argument also holds to explain the reopening of PNIPAm at high methanol concentrations.
Further suggesting that the bridging scenario seems to be the only possible
explanation to co-non-solvency \cite{mukherji13mac,mukherji14natcom,mukherji15jcp} and pressure induced reopening presented in this work.

\subsection{Potential of mean force}

Finally we want to study the thermodynamic origin of this interesting conformational transition. For this purpose we have calculated
the potential of mean force (PMF) between solute and solvent components. The PMF is calculated using the umbrella sampling \cite{umbrella} over a
series of independent simulations at 298 K temperature and 500 MPa pressure, each for a $10~{\rm ns}$ long trajectory. The center-of-mass positions between the
NIPAm monomer and the solvent components are generated by pulling the solvent component towards the NIPAm monomer
using a steered molecular dynamics algorithm. Here the spring constant is chosen as $1000~{\rm kJ~mol^{-1} nm^{-2}}$
and a velocity of pull was selected as $0.001~{\rm nm~ps^{-1}}$. Between 0 and 1.65 nm we choose 120 positions that are constrained
using a LINCS algorithm \cite{lincs}. The PMF is calculated by integrating the constraining forces $f_c$ using the
expression \cite{sprikjcp98,kahlen14jpcb},
\begin{equation}
V_{\rm PMF} (r) = \int_{r_0}^r \left[ \left<f_{c}\right>_{s} + \frac {2k_{\rm B}T}{s} \right] ds + const..
\label{eq:pmf}
\end{equation}
Here $\left<f_{c}\right>_{s}$ is the average force at a distance $s$ between the NIPAm and respective solvent component.
$r_0$ represents the closest proximity that the solvent can approach a NIPAm monomer.
The factor $2k_{\rm B}T/s$ is the entropic correction. The constant term is taken such that the potential goes asymptotically to
zero at 1.4 nm.

\begin{figure}[ptb]
\begin{center}
\includegraphics[width=0.46\textwidth,angle=0]{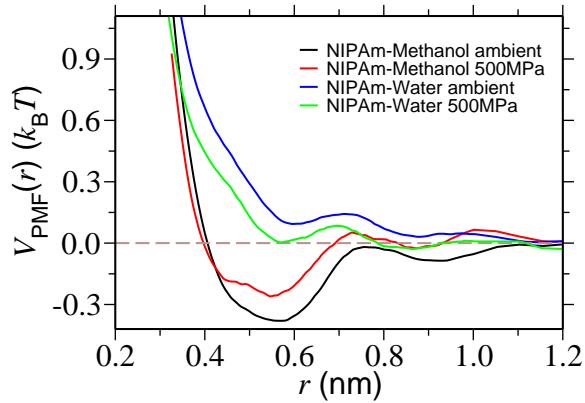}
\end{center}
\caption{Potential of mean force $V_{\rm PMF} (r)$ showing NIPAm-methanol and NIPAm-water interaction strengths
for two different pressures. Simulations are performed at a temperature of 298 K.
\label{fig:pmf}}
\end{figure}

In Fig.~\ref{fig:pmf} we show $V_{\rm PMF} (r)$.
Looking into the plot under ambient pressure, it becomes aparent that there exists an attractive well for NIPAm-methanol
interaction (represented by a black curve), whereas NIPAm-water interaction is repulsive (represented by a blue curve).
Furthermore, when the high pressure is applied
the attractive well of NIPAm-methanol interaction becomes shallower, indicating a reduced attractive interacting strength between
NIPAm and methanol at high pressure. On the other hand NIPAm-water develops a attractive well.
The applied pressure, therefore, could decrease the preferentiability of NIPAm-methanol interaction and, at the same time, enhancing
the NIPAm-water coordination, leading to polymer swelling.

\section{Conclusions}
\label{sec:conclu}

Using molecular dynamics simulations of an all atom model, we unveil the microscopic
origin why the application of
high hydrostatic pressures can destroy the co-non-solvency phenomenon of poly(N-isopropylacrylamide) (PNIPAm)
in aqueous methanol mixtures \cite{hofmann14pol}. Performing structural and thermodynamic analysis, we propose that the reopening of a
collapsed PNIPAm at 25\% methanol concentration is due to the partial loss of preferential binding of methanol with PNIPAm at
high pressures, which is the only key factor behind the polymer collapse in a mixture of two competing good solvents \cite{mukherji14natcom}.
This reduced preferentiability makes the background fluid reasonably homogeneous for  PNIPAm.
Thus is consistent with the swollen structure of the polymer under high pressures.
Additionally, the results presented here, eliminates any possible explanation of co-non-solvency effect based on pure entropic effects.
Had the collapse-swelling transition been dictated by depletion forces, polymer would have never open up under high pressures, especially
because depletion forces are most severe under high pressures.

\vspace*{.2in}

\noindent

{\bf Acknowledgment}

The development of this work would not have been possible without fruitful collaboration with Carlos Marques, which we take
this oppurtuinity to gratefully acknowledge. We thank Walter Richtering for bringing Ref. \cite{hofmann14pol} to our attention
and Davide Donadio and Torsten Stuehn for many stimulating discussions.
T.E.O. and P.A.N. acknowledges financial support from CNPq and CAPES from Brazilian Government and hospitality at the
Max-Planck Institut f\"ur Polymerforschung, where this work was performed.
We thank Robinson Cortes-Huerto and Carlos Marques for critical reading of the manuscript.
Simulation snapshots in this manuscript are rendered using VMD \cite{schulten}.


\vspace*{.2in}

\begin{thebibliography}{26}

\bibitem{schild91mac}
Schild, H. G; Muthukumar, M.; Tirrell, D. A.;
{\it Macromolecules} {\bf 1991} {\it 24}, 948.

\bibitem{zhang01prl}
Zhang, G.; Wu, C.;
{\it Phys. Rev. Lett.} {\bf 2001} {\it 86}, 822.

\bibitem{walter12jpcb}
Walter, J.; Sehrt, J.; Vrabec, J.; Hasse, H. J.;
{\it J. Phys. Chem. B} {\bf 2012} {\it 116}, 5251.

\bibitem{koj12jpsb}
Kojima, H.; Tanaka, F.; Scherzinger, C.; Richtering, W.;
{\it J Pol. Sci. B} {\bf 2012} {\it 51}, 1100.

\bibitem{tanaka08prl}
Tanaka, F.; Koga, T.; Winnik, F. M.;
{\it Phys. Rev. Lett.} {\bf 2008} {\it101}, 028302.

\bibitem{mukherji13mac}
Mukherji, D.; Kremer, K.;
{\it Macromolecules} {\bf 2013} {\it 46}, 9158.

\bibitem{mukherji14natcom}
Mukherji, D.; Marques, C. M.; Kremer, K.;
{\it Nat. Commun.} {\bf 2014} {\it 5}, 4882.

\bibitem{cohen10natmat}
Cohen-Stuart, M. A.; Huck, W. T. S.; Genzer, J.; M\"uller, M.; Ober, C.; Stamm, M.; Sukhorukov, G. B.;
Szleifer, I.; Tsukruk, V. V.; Urban, M.; Winnik, F.; Zauscher, S.; Luzinov, I.; Minko, M.;
{\it Nature Materials} {\bf 2010} {\it 9}, 101.

\bibitem{ward11poly}
Ward, M. A.; Georgiou, T. K.;
{\it Polymers} {\bf 2011} {\it 3}, 1215.

\bibitem{sissi14natcom}
de Beer, S.; Kutnyanszky, E.; Sch\"on, P. M.; Vancso, G. J.; M\"user, M. H.;
{\it Nat. Commun.} {\bf 2014} {\it 5}, 3781.

\bibitem{mukherji15jcp}
Mukherji, D.; Marques, C. M.; Stuehn, T. Kremer, K.;
{\it J. Chem. Phys.} {\bf 2015} {\it 142}, 114903.

\bibitem{hofmann14pol}
Hofmann, C. H.; Grobelny, S.; Erlkamp, M.; Winter, R.; Richtering, W.;
{\it Polymer} {\bf 2014} {\it 55}, 2000.

\bibitem{gro}
Pronk, S.; Pall, S.; Schulz, R.; Larsson, P.; Bjelkmar, P.; Apostolov, R.; Shirts, M. R.; Smith, J. C.; Kasson, P. M.; van der Spoel, D.; Hess, B.; Lindahl, E.;
{\it Bioinformatics} {\bf 2013} {\it 29}, 845.

\bibitem{groms}
van Gunsteren, W. F.; Billeter, S. R.; Eising, A. A.; H\"unenberger, P. H.; Kr\"uger, P.; Mark, A. E.; Scott, W. R. P.; Tironi, I. G.;
{\it Hochschulverlag AG an der ETH Z¨urich} (1996).

\bibitem{spce}
Berendsen, H. J. C.; Grigera, J. R.; Straatsma, T. P.;
{\it J. Phys. Chem.} {\bf 1987} {\it 91}, 6269.

\bibitem{per01jpca}
Pereira, J. C. G.; Catlow, C. R. A.; and Price, G. D.;
{\it J. Phys. Chem. A} {\bf 2001} {\it 105}, 1909.

\bibitem{berend}
Berendsen, H. J. C.; Postma, J. P. M.; van Gunsteren, W. F.; DiNola, A.; Haak, J. R.;
{\it J. Chem. Phys.} {\bf 1984} {\it 81}, 3684.

\bibitem{pme}
Essmann, U.; Perera, L.; Berkowitz, M. L.; Darden, T.; Lee, H.; Pedersen, L. G. A.;
{\it J. Chem. Phys.} {\bf 1995} {\it 103} 8577.

\bibitem{mukherji12jctclet}
Mukherji, D.; van der Vegt, N. F. A.; Kremer, K.; Delle Site, L.;
{\it J. Chem. Theory Comput.} {\bf 2012} {\it 8}, 375.

\bibitem{degennesbook}
P.-G. de Gennes,
{\it Scaling Concepts in Polymer Physics}
(Cornell University Press, London, 1979).

\bibitem{desclobook}
J. Des Cloizeaux and G. Jannink,
{\it Polymers in Solution: Their Modelling and Structure}
(Clarendon Press, Oxford, 1990).

\bibitem{net02}
Netz, P. A.; Starr, F. W.; Barbosa, M. C.; Stanley, H. E.;
{\it Physica A} {\bf 2002} {\it 314}, 470.

\bibitem{net04}
Netz, P. A.; Starr, F. W.; Barbosa, M. C.; Stanley, H. E.;
{\it Braz. J. Phys.} {\bf 2004} {\it 34}, 24.

\bibitem{kb51jcp}
Kirkwood, J. R.;  Buff, F. P.;
{\it J. Chem. Phys.} {\bf 1951} {\it 19}, 774.

\bibitem{rosgen05bio}
R\"osgen, J.; Pettitt, B. M.; Bolen, D. W.;
{\it Biophys. J} {\bf 2005} {\it 89}, 2988.

\bibitem{lekerbook}
H. N. W. Lekkerkerker and R. Tuinier,
{\it Colloids and the Depletion Interaction}
(Clarendon Press, Oxford, 1990).

\bibitem{umbrella}
Torrie, G. M.; Valleau, J. P.;
{\it J. Comp. Phys.} {\bf 1977} {\it 23}, 187.

\bibitem{lincs}
Hess, B.; Bekker, H.; Berendsen, H. J. C.; Fraaije, J. G. E. M.;
{\it J. Comp. Chem.} {\bf 1997} {\it 18}, 1463.

\bibitem{sprikjcp98}
Sprik, M.; Ciccoti, G.;
{\it J. Chem. Phys.} {\bf 1998} {\it 109}, 7737.

\bibitem{kahlen14jpcb}
Kahlen, J.; Salimi, L.; Sulpizi, M.; Peter, C.; and Donadio, D.
{\it J. Phys. Chem. B} {\bf 2014} {\it 118}, 3960.

\bibitem{schulten}
Humphrey, W.; Dalke, A.; Schulten, K.;
{\it J. Mol. Graph.} {\bf 1996} {\it 14}, 33.

\end{thebibliography}
\end{document}